\documentclass{article}
\usepackage{spconf,amsmath,graphicx}
\usepackage{mathtools}
\usepackage{enumitem}
\setlist{nosep, leftmargin=14pt}

\usepackage{mwe} 

\title{Handling Label Uncertainty on the Example of Automatic Detection of Shepherd's Crook RCA in Coronary CT Angiography}
\name{
    \begin{tabular}{c}
        Felix Denzinger$^{1, 2}$, Michael Wels$^{2}$, \textit{Oliver Taubmann}$^{2}$, Florian Kordon$^{1}$, Fabian Wagner$^{1}$,\\ 
        Stephanie Mehltretter$^{1} $, \textit{Mehmet A. G\"uls\"un}$^{2}$, \textit{Max Sch\"obinger}$^{2}$, \textit{Florian Andr\'e}$^{3}$, \textit{Sebastian Bu\ss}$^{3}$,\\ \textit{Johannes G\"orich}$^{3}$, \textit{Michael S\"uhling}$^{2}$, \textit{Andreas Maier}$^{1}$, \textit{Katharina Breininger}$^{4}$
    \end{tabular}
}
 \address{ 
     $^{1}$ Pattern Recognition Lab, FAU Erlangen-N\"urnberg, Erlangen, Germany\\
     $^{2}$ Siemens Healthcare GmbH, Computed Tomography, Forchheim, Germany\\
     $^{3}$ Das Radiologische Zentrum,
     Sinsheim-Eberbach-Erbach-Walldorf-Heidelberg, Germany\\
     $^{4}$ Department Artificial Intelligence in Biomedical Engineering, FAU Erlangen-Nürnberg
     }
    
\begin{document}
%
\maketitle
\begin{abstract}
Coronary artery disease (CAD) is often treated minimally invasively with a catheter being inserted into the diseased coronary vessel. If a patient exhibits a Shepherd's Crook (SC) Right Coronary Artery (RCA) -- an anatomical norm variant of the coronary vasculature -- the complexity of this procedure is increased. Automated reporting of this variant from coronary CT angiography screening would ease prior risk assessment. 
We propose a 1D convolutional neural network which leverages a sequence of residual dilated convolutions to automatically determine this norm variant from a prior extracted vessel centerline. As the SC RCA is not clearly defined with respect to concrete measurements, labeling also includes qualitative aspects. Therefore, 4.23\,\% samples in our dataset of 519 RCA centerlines were labeled as unsure SC RCAs, with 5.97\,\% being labeled as sure SC RCAs. We explore measures to handle this label uncertainty, namely global/model-wise random assignment, exclusion, and soft label assignment. Furthermore, we evaluate how this uncertainty can be leveraged for the determination of a rejection class. With our best configuration, we reach an area under the receiver operating characteristic curve (AUC) of 0.938 on confident labels. Moreover, we observe an increase of up to 0.020 AUC when rejecting 10\,\% of the data and leveraging the labeling uncertainty information in the exclusion process.

\end{abstract}
\section{Introduction}
\label{sec:intro}

\begin{figure*}
    \centering
    \includegraphics[width=\textwidth]{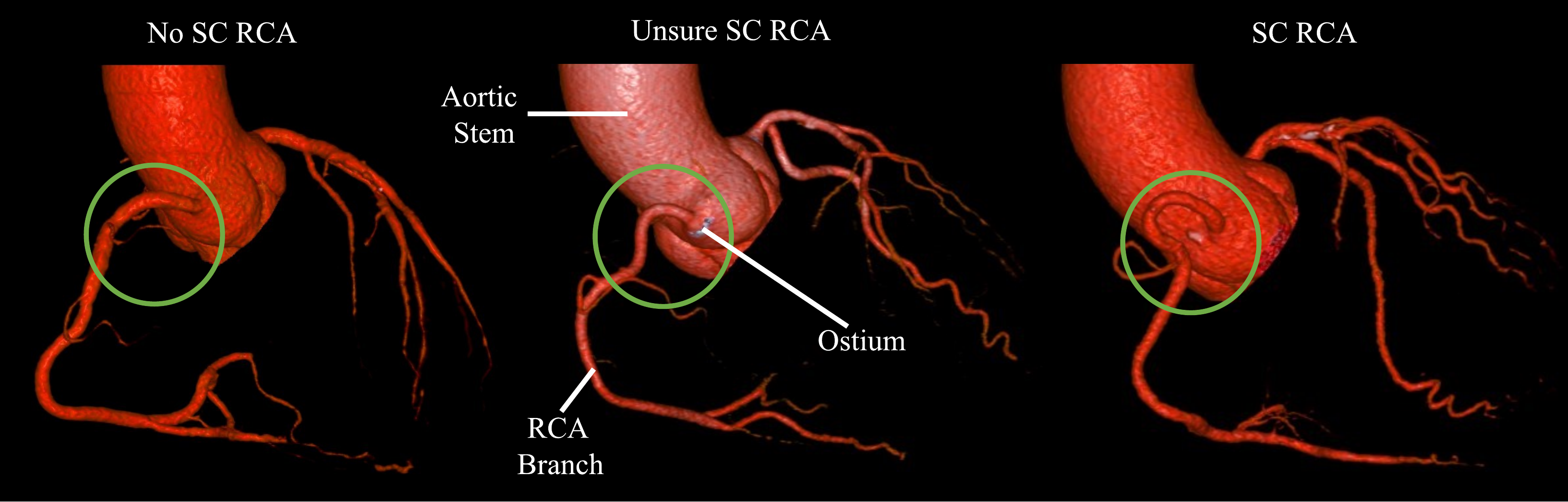}
    \caption{Volume rendering of the aortic stem and coronary arteries of three different patients: left) patient without a Shepherd's Crook (SC) RCA, center) patient labeled as having an unsure SC RCA as the RCA does take a tortuous high turn but to a lesser extent, and right) patient with a SC RCA defined by a high, tortuous turn after the origin of the RCA segment.}
    \label{fig:renderings}
\end{figure*}

Coronary artery disease (CAD) is an often deadly disease commonly linked to atherosclerotic plaque deposits narrowing the coronary vasculature~\cite{fuster92}. These lesions are usually treated minimally invasively in the cath lab, where a catheter is inserted through the femoral artery. This catheter is then guided toward the location of the lesion. The vessel at the lesion is then widened and stabilized using a balloon and a stent. One anatomical norm variant of the coronary vasculature -- the Shepherd's Crook right coronary ascending artery (SC RCA) -- may complicate this procedure as the RCA branch takes a high and tortuous turn directly after the ostium (cf. Fig.~\ref{fig:renderings} right)~\cite{shriki12}. Furthermore, this variant is suspected to increase the risk of developing CAD~\cite{saglam15}. Therefore, automated detection of SC RCA, e.g., from coronary CT angiography (CCTA) scans, is of interest. However, to the best of our knowledge, no prior work on this topic exists.

To develop a deep learning-based algorithmic solution, we build a data collection of 519 patients with labels indicating whether patients exhibit an SC RCA or not. However, as the sole definition of this norm variant is the high, tortuous turn, we did not only identify 31 cases we consider sure SC RCAs but also 22 border cases which we labeled as unsure, with an example displayed in Fig.~\ref{fig:renderings}. As these cases could not be labeled with high confidence by human readers, a machine learning approach should also rather not report a prediction for such samples instead of confidently predicting a label. Therefore, methods from uncertainty estimation or abstention learning are considered, where instead of just learning to distinguish presence from absence, a rejection class is additionally determined from the output of a machine learning model.

In summary, we formulate the following research questions:
\begin{enumerate}
    \item Can we automatically determine whether a patient has a SC RCA using a data-driven algorithm?
    \item How should samples be handled for which the annotator is not confident?
    \item Can we leverage the labeling uncertainty to enhance or at least better evaluate learning with abstention?
\end{enumerate}
We tackle them with the following contributions:
\begin{enumerate}
    \item Development of a deep learning approach which analyzes the centerline course of the RCA using a WaveNet-like 1D convolutional neural network. 
    \item Analysis of four different ways to handle the cases labeled as unsure: exclusion during training, randomly assigning a class either globally or for each model in an ensemble, or assigning a soft label.
    \item Proposal of a non-invasive percentile-based rejection scheme and examination of whether information about the frequency of uncertain samples can improve it. 
\end{enumerate}

\section{Material and Methods}
\subsection{Data}
Within this study, a data collection of 519 CCTA scans is used. Of these, 31 (5.97\,\%) are labeled as positive SC RCA cases, and 22 (4.23\,\%) as unsure. Labeling was performed by a doctoral researcher with four years of experience in the field of CAD assessment from CCTA scans. Centerlines of these scans were extracted using the well-established and robust algorithm of Zheng et al.~\cite{zheng13}. From these, the first 64\,mm (256 points, spacing of 0.25\,mm) of the RCA were extracted by combining the proximal and middle RCA segments as provided by the labeling algorithm of Denzinger et al.~\cite{denzinger22}. We consider the coordinates of the centerline as features that are normalized by subtracting the coordinates of the first centerline point (ostium) from all points and then dividing by 64\,mm, as this is the maximal possible length of an input centerline.

\subsection{Deep Learning-based Shepherd's Crook Detection}\label{sec:method}
\begin{figure}
    \centering
    \includegraphics[width=0.38\textwidth]{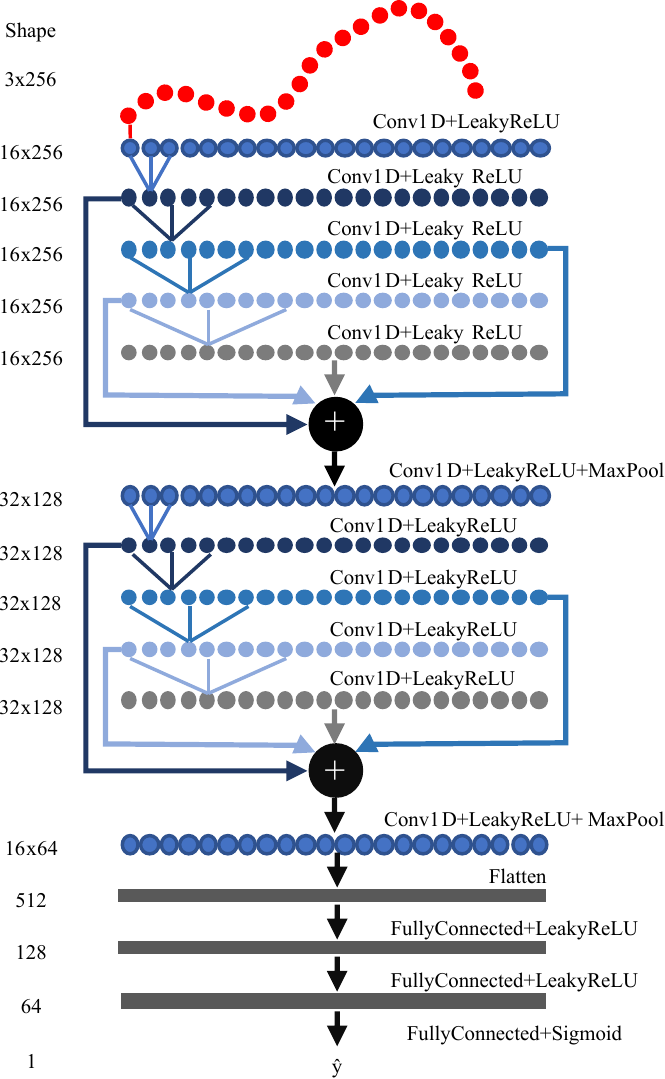}
    \caption{Overview of the WaveNet-inspired 1D convolutional neural network for the classification of SC RCA. The input to the network are the 3D centerline coordinates, which are processed by a set of 1D convolutional layers with increasing dilation grade. The features created from different perceptive fields are summed up and fed into a second WaveNet-like block. The final feature representation is then processed by a multi-layer perceptron to predict the presence of SC RCA.}
    \label{fig:model}
\end{figure}
As the local and global curvature and the overall course of the centerline are key features to be determined by a classifier, we propose to use a WaveNet-inspired~\cite{wavenet} deep learning architecture as depicted in Fig.~\ref{fig:model}. It leverages 1D convolutions of differing dilation grades to model short and long-range dependencies, which are combined and weighted by a multi-layer perceptron~\cite{maier2019gentle}.

To prevent overfitting, we randomly rotate our training data with the ostium as the rotation center in a range of up to 45\,$^{\circ}$ in all directions.
Furthermore, we use a binary cross-entropy loss, an Adam optimizer with the default learning rate of 0.001, and default batch size of 32. At test time, data is augmented by rotations of $[-15\,^{\circ},0\,^{\circ},15\,^{\circ}]$, with the final prediction being the mean across all rotations.
We use a fixed amount of 100 epochs to omit the need for a validation set due to the small number of SA RCA in the dataset. To improve the robustness of the prediction under this setting, we combine five training runs to form one final model by averaging the predictions of the five sub-models.
Due to the limited amount of data, the performance statistics differed for repeated experiments. To obtain reproducible results, we performed a 5-fold cross-validation and repeated it 25 times. Data was split in a stratified manner regarding both positive and unsure samples.

\subsection{Label Uncertainty Handling}
In this work, we evaluate four different strategies to handle samples with unsure labels: randomly assigning them to one class globally (``Fixed'') or for each training run (``Varied''), not including them in the training phase (``Exclusion''), or assigning a soft label of 0.5 (``0.5'').

The global random assignment of all samples mimics the usual handling of unsure cases, which are, in practice, not labeled as such, but some class assignment is enforced. With the random assignment for each individual training run of the ensemble and then combining the prediction over 5 of these runs, the output probabilities for the unsure cases should lie in between the distributions of the sure cases. 

Not including the unsure samples is also a valid strategy but decreases the amount of data seen by the network.

Assigning a soft label of 0.5 to the unsure samples encourages the probabilities of the network to form a separate distribution between the negative and positive classes.

\subsection{Percentile-based Abstention}
As there are samples marked as unsure, evaluating whether this labeling uncertainty can be confirmed by model uncertainty is an obvious choice. 
Therefore, we perform learning with abstention, i.e., determining which samples should be rejected. Because of the class imbalance, defining an abstention rule around the probability value of 0.5 (anchor) is not applicable. 
Instead, we propose a percentile-based approach: a frequently reported value for the prevalence $p(y_1)$ of the SC RCA is 5\,\%\cite{gossman1988percutaneous}. Therefore, for an ideal classifier the highest 5\,\% of the test-set predictions would belong to the positive class. To account for this, we run inference on the entire test set and select the prediction value at the 95th percentile (1-$p(y_1)$) in the probability histogram as the anchor of our abstention interval. 

From our defined anchor, we define our exclusion interval as:
\begin{equation}
    \begin{split}
        p_\textrm{min} &= 1 - p(y_1) - e * p(y_0)\\
        p_\textrm{max} &= 1 - p(y_1) + e * p(y_1)
    \end{split}
\end{equation}
with $p_\textrm{min}$ and $p_\textrm{max}$ also referring to percentile values in the histogram and $e\in[0;1]$ denoting the exclusion rate, which can be varied to specify the amount of coverage, i.e., the amount of data kept after abstention. With this interval, we keep the class balance also after abstention, as samples are excluded in relation to the prior probability of both classes. Note that the probability values corresponding to the percentiles can be transferred to new single samples as well.

Additionally, we examine whether a better abstention interval can be achieved by additionally leveraging the information of the frequency of the unsure samples observed in the training data. To this end, we replace $p(y_1)$ with $p(y_1)+p(y_{0.5})$ and $p(y_{0})$ with $p(y_0)-(p(y_1)+p(y_{0.5}))$ in the interval defined above. We call this configuration ``$p(y_{0.5})$''. 

\subsection{Evaluation}
Since there are no ground truth labels for the unsure samples, we propose using the following three performance measures: we calculate the AUC for all possible permutations of class assignment for the unsure cases to get the best and worst possible AUC value. Additionally, we report the performance solely on the sure samples. 
As discussed in Section~\ref{sec:method}, 5-fold cross validation with 25 repetitions was used to obtain robust results given the small number of overall samples.

\section{Results}

Our results for the different evaluated configurations are displayed in Table~\ref{tab:results}. 
From a high-level perspective, there is a relatively large gap between the best and worst possible AUC, indicating how much of an impact the relatively small number of unsure cases can have during test time. Generally, we reach excellent performance on the data set consisting of high confidence labels with an AUC of up to 0.940 at 100\,\% coverage.

The choice of how to handle unsure cases during training had a small effect. There is a clear trend that the exclusion of the borderline cases leads to a worse performance. Having a random assignment for each single training run or globally performed comparably, with a soft label of 0.5 performing best.

Regarding abstention, one can recognize that the metrics stayed mostly the same when only excluding 5\,\% of the data and increased slightly at 10\,\%. Paired with the observation that the distance between the best and worst possible AUC is not decreasing, it becomes apparent that only a limited amount of unsure samples lies in this initial exclusion interval. 

However, when including the information about the frequency of both true positive and unsure cases ($p(y_{0.5})$) in the determination of the abstention interval, we notice an improvement of up to 0.021 for the worst possible AUC at a coverage of 90\,\%. Also, the distance between the best and worst possible AUC decreases to a great extent, especially at a coverage of 75\,\%. This indicates that a majority of the unsure samples lies within this exclusion interval and therefore form a distinct distribution in the probability space. 

\begin{table}
    \centering
    \begin{tabular}{llllll}
Config&AUC&100\,\%&95\,\%&90\,\%&75\,\%\\
\hline

Exclusion&Best&0.942&0.942&0.946&0.963\\
Fixed&Best&0.944$^*$&0.946&0.951&\textbf{0.971}\\
Varied&Best&0.944&0.945&0.950&0.969\\
0.5&Best&\textbf{0.945}&0.947&0.954&\textbf{0.971}\\
0.5~$p(y_{0.5})$&Best&\textbf{0.945}&\textbf{0.951}&\textbf{0.958}&0.970\\
\hline
Exclusion&Worst&0.878&0.874&0.877&0.922\\
Fixed&Worst&0.885$^{***}$&0.885&0.890&0.939\\
Varied&Worst&0.885&0.885&0.892&0.940\\
0.5&Worst&\textbf{0.887}&0.887&0.894&0.941\\
0.5~$p(y_{0.5})$&Worst&\textbf{0.887}&\textbf{0.899}&\textbf{0.914}&\textbf{0.950}\\
\hline
Exclusion&Sure&0.934&0.931&0.933&0.954\\
Fixed&Sure&0.937$^{**}$&0.937&0.941&0.965\\
Varied&Sure&0.937&0.937&0.941&0.964\\
0.5&Sure&\textbf{0.938}$^*$&0.939&0.945&0.966\\
0.5~$p(y_{0.5})$&Sure&\textbf{0.938}&\textbf{0.945}&\textbf{0.953}&\textbf{0.967}\\

    \end{tabular}
    \caption{Performance with respect to the AUC for different handling of unsure cases with differing amount of coverage. ``Best'' and ``Worst'' are determined by calculating the AUC for all possible label assignments of the unsure samples. The ``Sure'' AUC is calculated only using the samples labeled with high confidence. Note that 0.5 and 0.5~$p(y_{0.5})$ refer to the same training configuration with different abstention parameterization. Significance was determined using paired-sample t-test. For 100\,\% coverage, the significance levels are displayed as $^{*}$ in relation to the next worst configuration with respect to the AUC with the following p-value thresholds: $^{*} \coloneqq p<.05$, $^{**} \coloneqq p<0.01$ and $^{***} \coloneqq p<0.002$ $p^*\coloneqq0.002$ according to Bonferroni correction.  }
    \label{tab:results}
\end{table}
\section{Discussion}

First, we want to discuss the model design choice.  We are unaware of any work performing classification regarding the course of the centerline. However, there are related works on registration or segment labeling of centerlines which either utilize 1D convolutions without dilation~\cite{car} or recurrent neural networks~\cite{wu19,fischer20}. We tested these approaches in initial experiments, but the model proposed in this manuscript yielded better results. Therefore, this WaveNet-like feature extraction might also be applicable to other approaches in this area.

Architectures like PointNet~\cite{pointnet} or ones based on graph deep learning~\cite{gdl} are more complex alternatives for the task at hand. These approaches might face additional challenges due to overfitting on a global structure or struggle to learn the internal connectivity of the centerline but this could be explored in future work.

Regarding our second research question of how to handle unsure samples, two related research fields come into mind: combating label noise and how to merge multiple annotators. 
Methods to combat label noise might be able to improve results presented in this paper. However, these algorithms do not answer the question on how to handle the unsure cases in the first place. With multiple annotators, strategies like taking the majority vote as confident label or ensembles, for which every sub-model is trained on a different annotator exist. These concepts are very similar to the strategies we evaluated here. A similar concept for introducing soft labels from multiple annotators was proposed very recently~\cite{softLabel22}. However, there are no works linking this concept to abstention yet. 

Regarding abstention, we propose a strategy that can be directly applied to a trained model. There are other methods for uncertainty estimation and abstention which either try to estimate an underlying gamma distribution in the probability space for each sample~\cite{ghesu19} or include a dedicated abstention class~\cite{thulasidasan19}. However, these are usually more invasive in that it needs more adaptations and a trained model cannot be taken as is.
In contrast, the abstention strategy we propose is simple and non-invasive. The most similar approach we found directly predicts uncertainty as a side-task and performs percentile-based abstention based on this uncertainty output~\cite{barnes21}.

\section{Conclusion}
We tackled three research questions in our work: can we automatically detect SC RCA, how should unsure samples be handled and does the information on how many of these samples exist help to form a rejection class?

An affirmative answer to the first question is given by the strong performance of 0.938 AUC on the confidently labeled data.
For the second research question we evaluated a set of strategies and found small differences in the performance yield. Overall, soft label assignment performed best. We hope that our work inspires others to perform similar analyses, especially as the concept of uncertain labels for single annotators is currently not widely applied in the medical domain.
Finally, we proposed a percentile-based abstention strategy. Here, we showed that adding information regarding the frequency of unsure cases improved the performance by a large margin for different levels of coverage. 

\section{Compliance with ethical standards}
\label{sec:ethics}

The CT examinations were clinically indicated by the referring physicians and conducted in accordance with current clinical standards, guidelines, and recommendations. The study was performed in accordance with the Declaration of Helsinki and was approved by the local ethics committee (S-226/2016 and S-758/2018, Ethikkommission der Medizinischen Fakult\"at Heidelberg, Germany). Subjects included as of January 2019 gave informed consent in the scientific data analyses. For the retrospective analyses of the datasets acquired before January 2019, a waiver of consent was granted by the aforementioned ethics committee.

\section{Acknowledgments}
\label{sec:acknowledgments}
This work was partially funded by Siemens Healtcare GmbH, Erlangen, Germany.

K.B.~gratefully acknowledges the support of d.hip campus - bavarian aim in form of a faculty endowment.

\bibliographystyle{IEEEbib}
\bibliography{strings,refs}

\end{document}